\documentclass[12pt]{iopart}
\usepackage{graphicx, epsfig}

\newcommand{\apjl}{{Astrophys.~J.~Lett.}}

\def\fun#1#2{\lower3.6pt\vbox{\baselineskip0pt\lineskip.9pt
        \ialign{$\mathsurround=0pt#1\hfill##\hfil$\crcr#2\crcr\sim\crcr}}}

\def\dslash{\not{\hbox{\kern-2pt $\partial$}}}
\def\Dslash{\not{\hbox{\kern-4pt $D$}}}
\def\Oslash{\not{\hbox{\kern-4pt $O$}}}
\def\Qslash{\not{\hbox{\kern-4pt $Q$}}}
\def\pslash{\not{\hbox{\kern-2.3pt $p$}}}
\def\kslash{\not{\hbox{\kern-2.3pt $k$}}}
\def\qslash{\not{\hbox{\kern-2.3pt $q$}}}

 \newtoks\slashfraction
 \slashfraction={.13}
 \def\slash#1{\setbox0\hbox{$ #1 $}
 \setbox0\hbox to \the\slashfraction\wd0{\hss \box0}/\box0 }

\def\ee{\end{equation}}
\def\be{\begin{equation}}

\begin{document}
\title{The Cosmic Neutrino Background and the age of the Universe} 
\author{Francesco de Bernardis$^1$, Alessandro
  Melchiorri$^1$, Licia Verde$^{2,3}$ and  Raul Jimenez$^{2,3}$\\
$^1$Physics Department and Sezione INFN, University of Rome ``La Sapienza'', P.le Aldo Moro 2, 00185 Rome, Italy\\
$^2$ ICREA \& Institute of Space Sciences (CSIC-IEEC), Campus UAB, Bellaterra, Spain.\\
$^3$Dept. of Astrophysical Sciences, Princeton University, Ivy lane, Princeton, NJ 08544, USA.\\
}

\begin{abstract}

We discuss the cosmological degeneracy between the age of the
Universe, the Hubble parameter 
and the effective number of relativistic particles $N_{\rm eff}$. 
We show that independent determinations of the Hubble parameter $H(z)$ such 
as those recently provided by \cite{svj}, combined with other cosmological data sets can provide the most stringent
 constraints on $N_{\rm eff}$, yielding $N_{\rm eff}=3.7_{-1.2}^{+1.1}$
 at $95 \%$ confidence level. A neutrino background is detected  with
 high significance: $N_{\rm eff}>1.8$ at  better than 99\% confidence
 level. Constraints on the age of the Universe in the framework
of an extra background of relativistic particles are improved by a factor
$3$.
\end{abstract}



\maketitle

\section{Introduction}
The recent measurements of cosmic microwave background (CMB) anisotropies and 
polarization \cite{spergelwmap1, spergelwmap2}, alone or in combination with other cosmological data sets,  have provided  confirmation of the standard
cosmological model and an accurate determation 
of some of its key parameters. 
In particular, the new determination
of the age of the Universe $13.8 \pm 0.3$ Gyrs improves by an
order of magnitude previous determinations from, e.g.,
cosmochronology  of long-lived radioactive nuclei \cite{thorium} and 
population synthesis of the oldest stellar populations \cite{Jimenez96,dunlop96,chaboyerkrauss03}.

It is important however to constrain possible deviations from the standard cosmological model.
Here we concentrate on constraints  on the amount of relativistic energy density at recombination:  cosmological results can be dramatically affected if  assumptions about the physical energy density in relativistic particles $\omega_{\rm rel}$ are relaxed.   The shape of the CMB angular power spectrum is sensitive to the epoch of matter-radiation equality:  a change in $\omega_{\rm rel}$ can be compensated by a change in the physical  cold dark matter density $\omega_c$,  in the Hubble constant $H_0$  and, to less extent,  in the power spectrum spectral slope $n_s$ \cite{bowen}.  Small scales observables (such as the CMB damping tail and large-scale structure) are also affected e.g., \cite{hallolivier, hannestad05, Pierpaoli03,basniskyseljack}.

In the standard model, $\omega_{\rm rel}$ includes photons and neutrinos,
and is often parameterized in terms of the equivalent number of standard massless neutrinos species $N_{\rm eff}$; in particular,
$\omega_{\rm rel} = \omega_{\gamma} + N_{\rm eff} \omega_{\nu}$,
where $\omega_{\gamma}$ is the energy density in photons and 
$\omega_{\nu}$ is the energy density in one active neutrino.  Measuring
$\omega_{\rm rel}$ thus gives a direct observation on the effective number
of neutrinos, $N_{\rm eff}$.
The standard model predicts three neutrino species; corrections to account for  QED effects and for neutrinos being not completely decoupled during electron-positron annihilation imply  $N^{SM}_{\rm eff}=3.04$ \cite{corrections1, corrections2, corrections3, corrections4, corrections5}. Any light particle that does not couple to electrons, ions and photons will act as 
an additional relativistic species.

Departures from the standard model which are described by a deviation
$N_{\rm eff}\neq 3$ can arise from  the decay of dark matter
particles \cite{bonometto98, lopez98,  hannestad98, kaplinghat01}, 
quintessence \cite{bean}, exotic models \cite{unparticles}, 
and additional hypothetical
relativistic particles such as a light majoron or a sterile neutrino.
Such hypothetical particles are strongly constrained from standard big
bang nucleosynthesis (BBN), where the allowed extra relativistic
degrees of freedom are  $N_{\rm eff}^{\rm BBN}=3.1_{+1.4}^{-1.2}$ (see e.g. \cite{mangano}).
When comparing $N_{\rm eff}$ constraints from BBN and CMB, one should keep in mind that  
 they rely on  different physics and  the energy density in
relativistic species may easily change from the
time of BBN ($T \sim$ MeV) to the time of last rescattering ($T \sim$
eV) in several non-standard models. Moreover,  the two estimates are affected by different systematics 
 (see \cite{mangano}).

Cosmological data analyses with variations in $N_{\rm eff}$
have been recently undertaken by many authors \cite{hannestad, hansen, Pierpaoli03, Peacock, Ichikawa1, Ichikawa2, crotty, elgaroy, barger, hannestad05,mangano,hamann}.
Here, we first analyze the degeneracy of cosmological
parameters  and in particular the age of the Universe, with $N_{\rm eff}$, 
and then show how determinations of the Hubble parameter can help break this degeneracy. 
We show that the recent determinations of $H_0$ from the
HST key project  \cite{hstkey} and $H(z)$ 
provided by  \cite{svj} (SVJ; based on  
\cite{data} and references therein) can provide, when combined 
with CMB and other cosmological data, new bounds on $N_{\rm eff}$. 

\section{Data Analysis: Method and Results}
The method we adopt is based on the publicly available Markov Chain Monte Carlo
package \texttt{cosmomc} \cite{Lewis:2002ah}. We  start by considering  a standard LCDM model described by the following  set of cosmological parameters, adopting flat priors on them:
the physical baryon and CDM densities, $\omega_b=\Omega_bh^2$ and 
$\omega_c=\Omega_ch^2$, the ratio of the sound horizon to the angular diameter
distance at decoupling, $\theta_s$, the scalar spectral index, $n_{s}$ and amplitude $A_s$,
and the optical depth to reionization, $\tau$.  For all these parameters the chosen boundaries of the priors do not affect the cosmological constraints.  To these parameters we add the 
possibility of having an extra-background of relativistic particles (parametrized 
by $N_{\rm eff} \neq 3.04$, see e.g. \cite{bowen}, with a flat prior  $0\le N_{\rm eff} \le15$)\footnote{While it is common to assume a flat prior on $\theta_s$, the standard version of  \texttt{cosmomc} uses a fitting formula to convert between $\theta_s$ and $H$ which is valid for $N_{rm eff}\equiv 3.04$. Thus running the standard  \texttt{cosmomc} with a flat prior on $\theta_s$ and $N_{\rm eff} \neq 3.04$ is equivalent to using a distorted prior on $H$. We find that this does not affect significantly the cosmological constraints, in particular for $50<H_0<80$ the difference is completely negligible.}. Later on, when we consider the possibility of a dark energy equation of state $w \neq  -1$, we assume a flat prior on $w$.
 Temperature, cross polarization and
  polarization CMB fluctuations from the WMAP 3 year data 
\cite{spergelwmap1,spergelwmap2,Page:2006hz,Hinshaw:2006ia} are considered.
WMAP data are combined  with the following data sets:  higher
resolution CMB experiments BOOMERanG, ACBAR, CBI  
and VSA \cite{wmapext1,wmapext2,wmapext3,wmapext4,wmapext5,wmapext6,wmapext7,wmapext8,wmapext9}, the power spectrum of
galaxies from the Sloan Digital Sky Survey (SDSS)   
\cite{2004ApJ...606..702T} and Two degree field  Galaxy Redshift Survey (2dFGRS)  \cite{2005MNRAS.362..505C} on linear scales ($k < 0.2$(Mpc/h)$^{-1}$); 
 the galaxy   bias $b$ is considered as an
additional nuisance parameter and is marginalized over.  Constraints
obtained from the supernova (SN-Ia) luminosity measurements are also included with data set from \cite{riess, 2006A&A...447...31A}. This combination  is referred to as "WMAP+ALL".

Finally, we consider Hubble key project determination of the Hubble constant \cite{hstkey} (HST) and  the SVJ \cite{svj}
determination of the redshift dependence of the Hubble parameter
$H(z)$ from observations of passively evolving galaxies (SVJ).

Since the SDSS and 2dFGRS data differ in the shapes of the two measured power spectra,  leading to a disagreement in their best fit values for $N_{\rm eff}$ \cite{spergelwmap1},  we also consider the reduced combination of WMAP alone and  WMAP+SDSS (this is the combination of CMB and large scale structure which give the weakest constraints in $N_{\rm eff}$). As we will show below, this reduced combination  with  the addition of   the Hubble parameter determinations provide  constraints  virtually indistinguishable form those obtained with the "ALL" data set.

When $\omega_{\rm rel}$  (or $N_{\rm eff}$) is  considered as a free parameter the constraint on the age 
of the Universe , $t_0$,  from cosmological data changes from
$13.84\pm0.28$ Gyrs to $13.8^{+2.3}_{-3.2}$ Gyrs:
there is a strong degeneracy between $\omega_{\rm rel}$ and $\omega_m$  which is mostly driven by a degeneracy between   $N_{\rm eff}$ and $H_0$.  
This is shown in the  first four columns of table 1 (for WMAP dataset alone and WMAP+ALL),  and the first two columns of table 3 (for WMAP+SDSS).

\begin{table}
\caption{\label{table2} Cosmological constraints ($95 \%$ c.l.) on the number of relativistic
particles and other cosmological parameters  from WMAP alone, WMAP+ALL, adding HST and SVJ data sets (see text for more details; a conservative prior $H_0 \le 100 Km/s/Mpc$ is included in the WMAP alone case)}
\begin{center}
\begin{tabular*}{0.9\textwidth}{@{\extracolsep{\fill}}|l||c|c|c|c|c|}
\hline
                    &   WMAP alone    & WMAP alone&    WMAP+ALL    & WMAP+ALL   \\
Parameter &$N_{\rm eff}=3.04$ & &$N_{\rm eff}=3.04$ &      \\

\hline
$N_{\rm eff}$  & $-$   &$5.5_{-3.9}^{+4.0}$  &  $-$      &$3.3_{-2.3}^{+4.0}$                \\
$H_0$&$73.2^{+3.1}_{-3.2}$        & $82^{+18}_{-16}$             &   $70.8^{+3.3}_{-3.1}$        & $71^{+21}_{-12}$              \\
$\omega_c$ &$0.1054^{+0.0078}_{-0.0077}$&$0.15^{+0.08}_{-0.07}$&$0.1087^{+0.0088}_{-0.0078}$&$0.113^{+0.086}_{-0.037}$\\
$\Omega_m$&$0.241^{+0.034}_{-0.034}$&$0.25^{+0.07}_{-0.06}$&$0.262^{+0.036}_{-0.032}$&$0.265^{+0.038}_{-0.039}$\\
age(Gyr) & $13.73^{+0.31}_{-0.30}$& $12.4^{+5.5}_{-6.0}$& $13.84^{+0.28}_{-0.28}$& $13.8^{+2.3}_{-3.2}$\\
\hline
\hline
\hline
                   &   WMAP+ALL   &   WMAP+ALL       &  WMAP+ALL& WMAP alone\\
Parameter &+HST $$& +SVJ $$ &  +HST+SVJ & +SVJ\\
\hline
$N_{\rm eff}$      &$3.03_{-1.7}^{+2.2}$           & $3.8^{+1.2}_{-1.15}$          & $3.71^{+1.17}_{-1.08}$&$4.0_{-1.2}^{+1.2}$\\
$H_0$  &$70.5^{+10.}_{-9.5}$           & $73.8^{+6.6}_{-6.5}$          & $73.5^{+6.1}_{-5.9}$ &$75.5^{+8.1}_{-7.4}$\\
$\omega_c$ &$0.109^{+0.045}_{0.029}$&$0.121^{+0.024}_{-0.021} $&$0.120^{+0.023}_{-0.020} $ &$0.124^{+0.035}_{-0.027}$\\
$\Omega_m$&$0.264^{+0.035}_{-0.036}$&$0.262^{+0.36}_{-0.040}$&$0.263^{+0.035}_{-0.036}$&$0.258^{+0.045}_{-0.058}$\\
age(Gyr) &$13.9^{+1.8}_{-2.0}$&$13.27^{+1.1}_{-0.97} $&$13.31^{+1.0}_{-0.91}$&$13.1^{+1.1}_{-1.1}$\\

%
\hline

\end{tabular*}
\end{center}
\end{table}
%
\begin{table}
\caption{\label{table3} As table 2 but for WMAP+SDSS. (see text for more details).}
\begin{center}
\begin{tabular*}{0.9\textwidth}{@{\extracolsep{\fill}} |l||c|c|c|c|c|c|}
\hline
                    &   WMAP+    & WMAP+& WMAP+   &   WMAP+       &  WMAP+\\
                    &  SDSS        & SDSS  & SDSS      &  SDSS          &    SDSS\\
& $N_{\rm eff}=3.04$   &                            &+HST $$             & +SVJ $$                    &  +HST+SVJ \\
\hline
$N_{\rm eff}$ & $-$ & $7.1^{+7.1}_{-5.1}$&$4.3^{+2.8}_{-1.6}$&$4.0^{+1.2}_{-1.1}$&$4.0^{+1.1}_{-1.1}$\\
$H_0$& $70.9^{+5.3}_{-4.7}$ &  $86^{+24}_{-21}$&$75^{+13}_{-21}$&$74.5^{+7.7}_{-7.3}$&$ 74.0^{+6.9}_{-6.6}$   \\
$\omega_c$ & $0.110^{+0.013}_{- 0.013}$& $0.18^{+0.13}_{-0.086} $&$0.132^{+0.068}_{-0.042}$&$0.128^{+0.027}_{-0.024}$&$ 0.127^{+0.027}_{-0.024}$\\
$\Omega_m$&$0.265^{+0.062}_{-0.056}$&$0.265^{+0.077}_{-0.065} $&$0.270^{+0.069}_{-0.060}$ &$0.271^{+0.065}_{-0.057} $ & $0.274^{+0.062}_{-0.055} $\\
age & $13.77^{+0.29}_{-0.30}$ &     $ 11.5^{+3.3}_{-2.7}$&$13.0^{+2.0}_{-2.0}$&$13.0^{+1.0}_{-0.92}$& $13.1^{+0.96}_{-0.87}$\\
\hline
\end{tabular*}
\end{center}
\end{table}
In  the first two columns of Table I we show the constraints from  WMAP alone:  including
variations in $N_{\rm eff}$ strongly affects the determination of most
of the parameters as the Hubble constant and the matter density. In
the last column of Table I, where we report the constraints obtained
from WMAP alone +the SVJ dataset. The constraints
on the age of the Universe are improved by a factor $3$ while
constraints on the Hubble parameter are now a factor $2$ better.
Age and $H(z)$ data are therefore extremely useful in breaking the
degeneracy. Consequently the  WMAP+SVJ analysis yields $N_{\rm eff}=4.0_{-1.2}^{+1.2}$.
 
\begin{figure}
\centering
\includegraphics[width=0.5\textwidth, angle=-90]{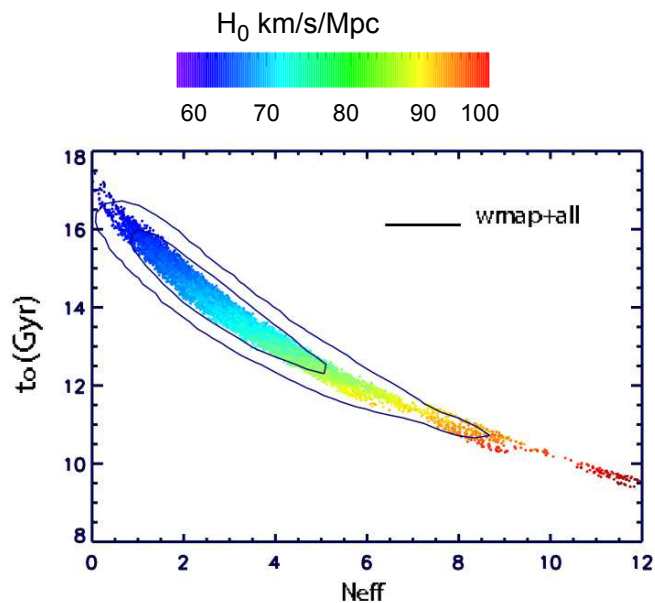}
\caption{Constraints on the $t_0-N_{\rm eff}$ plane
from current cosmological data (WMAP+ALL). A degeneracy between
the two parameters (and the Hubble constant) 
is evident. An extra background of relativistic particles with respect to the standard model prediction of $N_{\rm eff}=3.04$  is 
in agreement with current cosmological data  for ages of the Universe $t_0<13$ Gyr,  and Hubble constant  $H_0>70$km/s/Mpc.}
\label{age_neff}
\end{figure}

It is useful to examine the constraints from the cosmological dataset combination  WMAP+ALL 
on the $N_{\rm eff}-t_0$ plane  as shown in Fig. \ref{age_neff}. A strong degeneracy is present  and  the relativistic background is poorly constrained:  values of $N_{\rm eff}$ as
large as $8$ are  allowed at  $95\%$ c.l. and  letting $N_{\rm eff}$  vary
results in much weaker constraints on the current age of
the Universe $t_0$. As we can see, even if one includes most
of the current cosmological datasets, except  $H(z)$,  $N_{\rm eff}$ and 
$t_0$are poorly constrained.

\begin{figure}[htbp]
\centering
\includegraphics[width=0.7\textwidth]{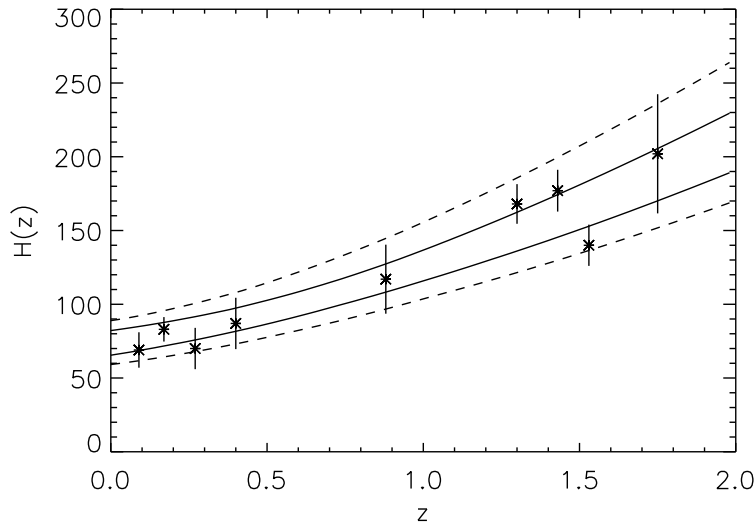}
\caption{The dashed lines show the range  of expansion histories, $H(z)$,  allowed at the 68\% c.l.  by the data set WMAP+SDSS +HST. The points with error-bars are the $H(z)$ determinations from \cite{svj}(SVJ)  and the solid line shows the range  of expansion histories allowed  at the 68\% c.l. after adding the constraints from these data.}
\label{Hz}
\end{figure}

Figure \ref{Hz} shows how $H(z)$ constraints reduce the parameter space allowed by the rest of the data combination while
table 2 (WMAP+ALL) and 3 (WMAP+SDSS)  show how the degeneracy can be broken and constraints on $N_{\rm eff}$ can be improved  by adding Hubble constant constraints; here HST denotes the HST Hubble key project $H_0$ determination \cite{hstkey} and SVJ denotes \cite{svj}.
The constraints obtained from the combination WMAP+ALL+HST+SVJ  are virtually indistinguishable from those obtained from the combination WMAP+SDSS+HST+SVJ. This implies that the tension between 2dFGRS and SDSS does not affect the result and that the  combination high-resolution CMB experiments+ 2dFGRS+Supernovae (SNLS \& GOLD) does not add a significant amount of information once the $H$ determinations are included.

\begin{figure}[htbp]
\includegraphics[width=3.0in,height=1.8in]{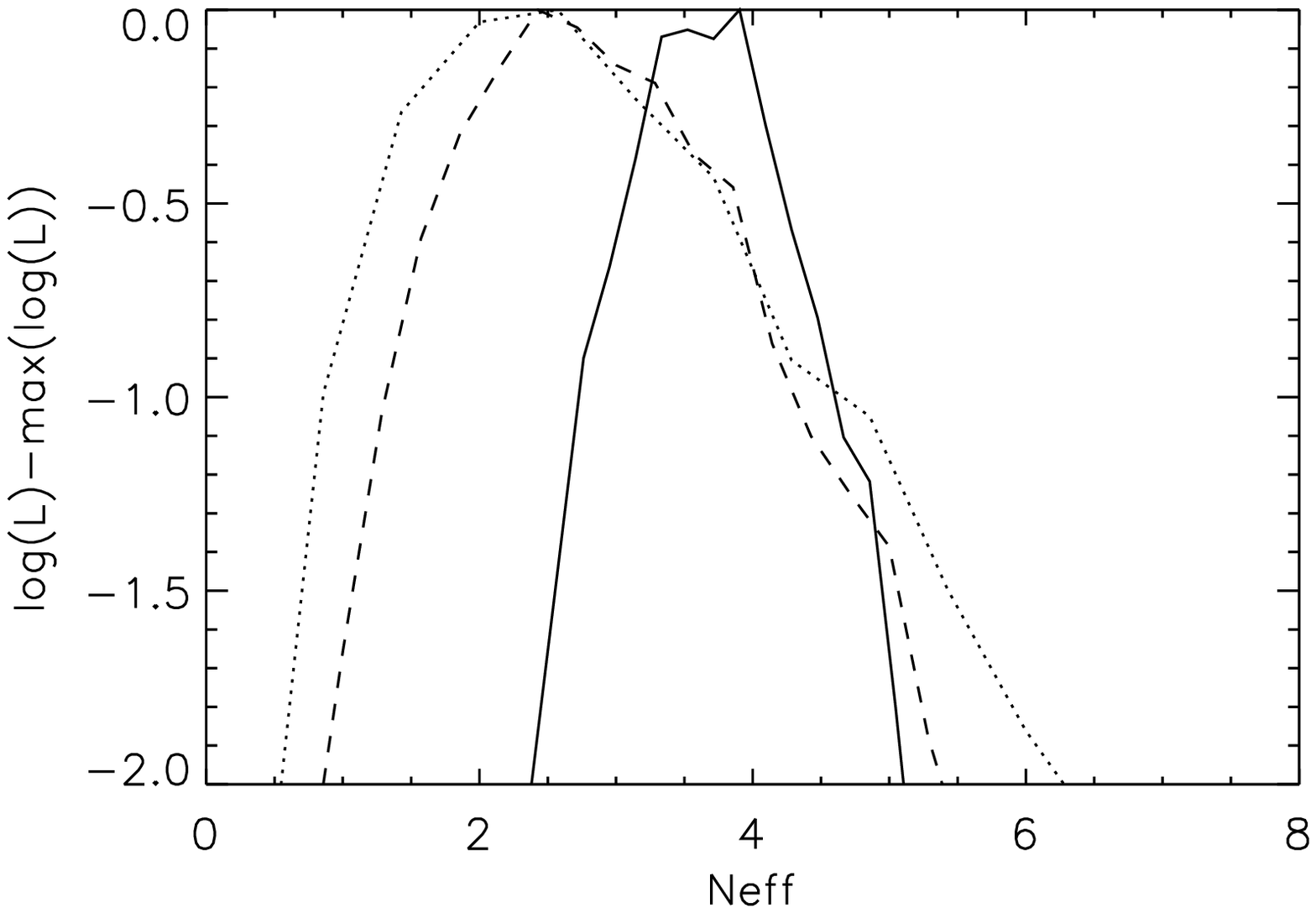}
\includegraphics[width=3.0in,height=1.8in]{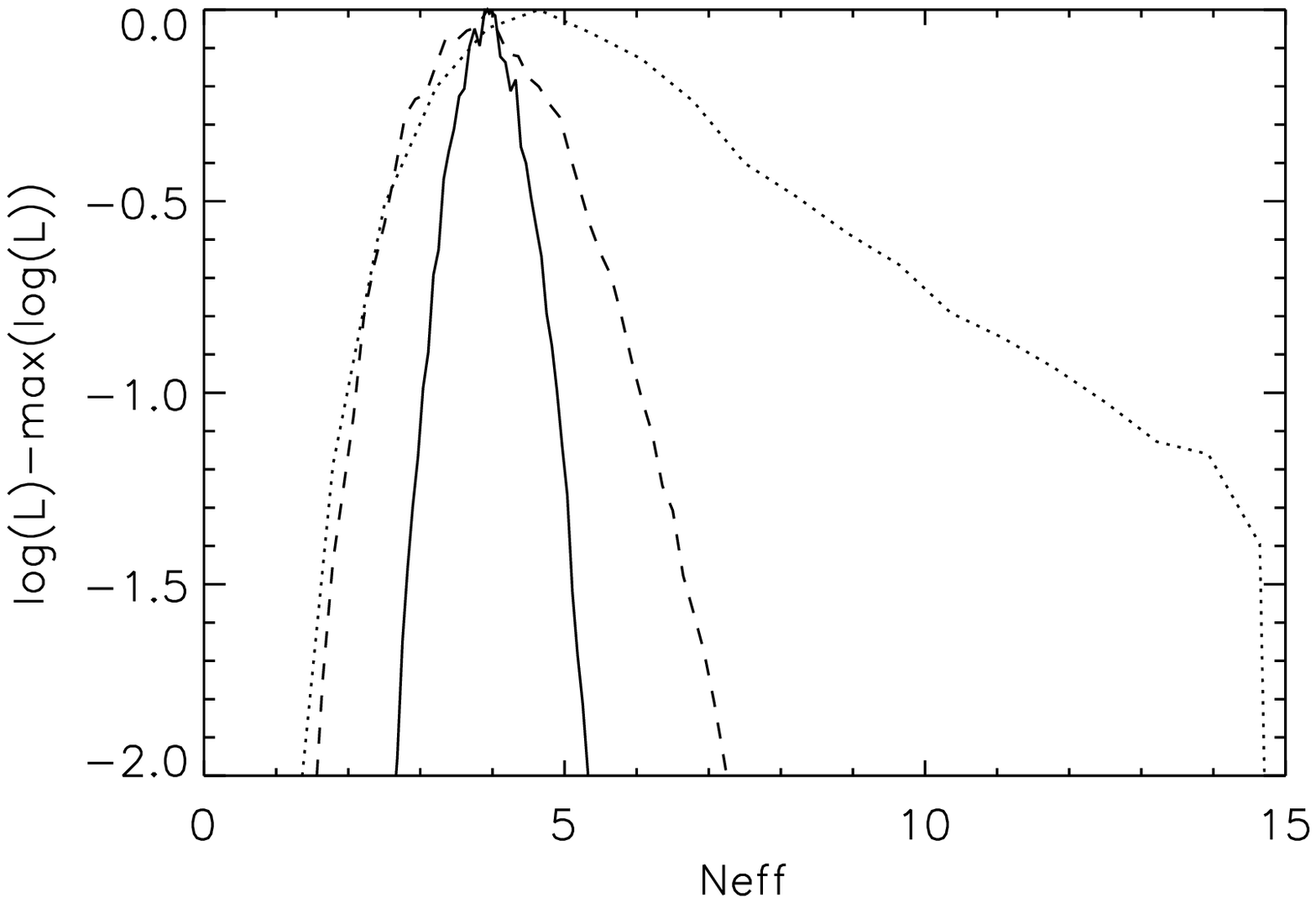}
\caption{Logarithm of the Likelihood ($\ln (L)$) as a function of $N_{\rm eff}$  marginalized over all other parameters and  normalized to be 0 at the maximum. Top: WMAP+ALL. Bottom:  WMAP+SDSS. Lines corresponds to the data set combination  without  $H$ determinations (dotted),  adding HST (dashed), adding  HST and SVJ (solid). }
\label{chi}
\end{figure}

We summarize the constraints on $N_{\rm eff}$ in figure \ref{chi}, where we show the likelihood ($L$)  as a function of $ N_{\rm eff}$ marginalized over all other parameters, for the case WMAP+ALL (top) and  WMAP+SDSS (bottom). Lines corresponds  to the data set combination  without  $H$ determinations (dotted),  adding HST (dashed) and  adding both HST and SVJ (solid). When  the Hubble parameter determinations are added  a neutrino background is detected at high significance: $N_{\rm eff}>1.8$ at  better than 99\% confidence level.

\begin{table}[htb]
\caption{\label{table4}Cosmological constraints ($95 \%$ c.l.) on the number of
  relativistic  particles allowing for a dark energy equation of
  state $w\neq-1$ from
  WMAP+2dF+SN-Ia and when adding the SVJ data.}
\begin{center}
\begin{tabular}{|l||c|c|c|c|c|c|}
\hline
                    &   WMAP+2dF    & WMAP+2dF\\
Parameter &+SN-Ia    &+SN-Ia+SVJ             \\
\hline
$N_{\rm eff}$ & $3.5_{-3.5}^{+3.8}$ & $3.8^{+1.1}_{-1.1}$\\
$w$&$-0.96_{-0.15}^{+0.14}$&$-0.95_{-0.16}^{+0.14}$\\
$H_0$& $72.7^{-13.2}_{+18.7}$ &  $74.2^{+6.8}_{-7.2}$\\
$\omega_c$ & $0.11^{+0.07}_{-0.04}$&$0.116_{-0.023}^{+0.027}$\\
$\Omega_m$&$0.25^{+0.04}_{-0.04}$&$0.251^{+0.042}_{-0.039} $\\
age(Gyr) & $13.7^{+2.5}_{-2.9}$ &     $ 13.2^{+1.1}_{-1.0}$\\
\hline
\end{tabular}
\end{center}
\end{table}

Finally, we consider the possibility of a dark energy equation of state $w\neq-1$. We include WMAP+2dF and SN-Ia datasets and we report the constraints in
table III.  The table shows that  allowing $w \neq-1$ produces little effect
on  $N_{\rm eff}$ when the SVJ data are included.

Very recently, several works have bounded the cosmological neutrino
background using cosmological data:  \cite{seljak} and \cite{mangano} 
report  $N_{\rm eff}= 4.6_{-1.6}^{+1.5}$. This limit, which is in
excellent agreement with the limit presented here,  
is obtained by including Lyman-$\alpha$ and baryonic acoustic
oscillation data, 
which are affected by completely different systematics from the 
Hubble parameter determinations used here (see \cite{hamann} for a discussion; systematics in the SVJ method are described  and quantified in \cite{data,dandrea}).

WMAP+SVJ (see Table I) combination which includes information
from $2$ datasets produces  $40 \%$ more
stringent constraints than those reported in these analyses which includes
more than $5$ datasets. 

\section{Conclusions}
We have considered how the Hubble parameter determinations of $H_0$ from  the HST key project \cite{hstkey} and of $H(z)$  from  passively evolving  galaxies \cite{svj}(SVJ) improve constraints on the physical energy density in relativistic particles, parameterized as an effective number of massless neutrino species $N_{\rm eff}$ which  can deviate from the standard model prediction of $N^{SM}_{\rm eff}=3.04$.  We find $N_{\rm eff}= 3.7^{+1.2}_{-1.1} $ (95\% c.l.)  from the dataset combination WMAP+ALL+HST+SVJ and  $N_{\rm eff}= 4.0^{+1.1}_{-1.1} $(95\% c.l.) from the reduced combination WMAP+SDSS+HST+SVJ. The WMAP+SDSS combination was chosen over e.g. WMAP+ other galaxy surveys, as SDSS  is the  large-scale structure data set  yielding the weakest constraints on $N_{\rm eff}$ when combined with WMAP.
These constraints are in good  agreement with the standard model prediction.
The addition of Hubble parameter determinations to the other cosmological data-sets break the degeneracy between $N_{\rm eff}$ and the age of the Universe.  
The nature of this degeneracy implies that there is not a range of redshift where $H(z)$ determinations would  yield the best constraints on $N_{\rm eff}$. On the other hand,  for a redshift range $0.8 < z < 2.0$ the corresponding age of the Universe is about $6.5$ to $3$ Gyr. At these ages and for solar metallicity, the UV spectra of galaxies ($2300 < \lambda/\AA < 4000$) are dominated by main sequence stars of F and  G class similar to our Sun\cite{Jimenez04}. This is a stage of stellar evolution that can be modeled accurately unlike stages in the post main-sequence dominated by mass-loss events. This will  therefore be the preferred redshift range to obtain further observations of passively evolving high-redshift galaxies to further constrain the number of relativistic species.

Future baryon acoustic oscillations spectroscopic  surveys will yield  very accurate measurements of $H(z)$, which can be used to further constrain  $N_{\rm eff}$. Forecasts for the  $N_{\rm eff}$ constraints from these surveys in combination with future CMB data,  will be presented  elsewhere.
%
\section*{Acknowledgments}

We acknowledge the use of the  Legacy Archive for Microwave Background Data Analysis (LAMBDA). Support for LAMBDA is provided by the  NASA Office of Space Science.
LV and RJ thank D. Spergel for discussions. We thank N. Wright  and S. Bashinsky for carefully reading the manuscript and for comments. 
%
\section*{References}
 \bibliographystyle{JHEP}
 \providecommand{\href}[2]{#2}\begingroup\raggedright
\end{document}